\documentclass{sig-alternate-05-2015}
\usepackage{multirow}
\usepackage{mathtools}
\usepackage{stmaryrd}
\usepackage{paralist}
\usepackage{url}
\usepackage{relsize}
\usepackage{multirow}
\usepackage{graphicx}
\usepackage{float}
\usepackage{multicol}
\usepackage{subfigure}
\usepackage{hyperref}
\usepackage{makeidx}  
\usepackage{graphicx}
\usepackage{pgfplots}
\usepackage{multirow}
\usepackage{amsmath}
\usepackage{amssymb}
\usepackage{subfigure}
\usepackage{ifsym}
\usepackage{stmaryrd}
\usepackage{algorithm}
\usepackage{algorithmic}
\usepackage{tikz,times}
\usepackage{xcolor}

\usepackage{lipsum}

\usepackage{tikz}
\usetikzlibrary{arrows,positioning,automata,decorations,fit,backgrounds,calc,shapes,snakes}
\usepgflibrary{shapes.geometric} 
\usetikzlibrary{shapes.geometric} 
\usepgflibrary{decorations.pathmorphing} 
\usetikzlibrary{decorations.pathmorphing} 
\usepackage{verbatim}
\usetikzlibrary{calc,backgrounds}
\usetikzlibrary{trees}

\begin{document}

\setcopyright{acmcopyright}

\doi{XX.XXX/XXXX}


\conferenceinfo{XXX}{XXX}


%

\title{hMDAP: A Hybrid Framework for Multi-paradigm Data Analytical Processing on Spark}
%
%
%
%
%
\numberofauthors{3} 
%
\author{
%
%
\alignauthor
Xiaowang Zhang\\
       \affaddr{School of Computer Science and Technology}\\
       \affaddr{Tianjin University, P. R. China}\\
       \affaddr{Tianjin Key Laboratory of Cognitive Computing and Application}\\
       \affaddr{Tianjin 300350, P.R. China}\\
       \email{xiaowangzhang@tju.edu.cn}
\alignauthor
Jiahui Zhang\\
       \affaddr{School of Computer Science and Technology}\\
       \affaddr{Tianjin University, P. R. China}\\
       \affaddr{Tianjin Key Laboratory of Cognitive Computing and Application}\\
       \affaddr{Tianjin 300350, P.R. China}\\
       \email{zhangjiahui@tju.edu.cn}
\alignauthor Zhiyong Feng\\
       \affaddr{School of Computer Software}\\
       \affaddr{Tianjin University, P. R. China}\\
       \affaddr{Tianjin Key Laboratory of Cognitive Computing and Application}\\
       \affaddr{Tianjin 300350, P.R. China}\\
       \email{zyfeng@tju.edu.cn}
       }
%

\maketitle
\begin{abstract}
We propose hMDAP, a hybrid framework for large-scale data analytical processing on Spark, to support multi-paradigm process (incl. OLAP, machine learning, and graph analysis etc.) in distributed environments. The framework features a three-layer data process module and a business process module which controls the former. We will demonstrate the strength of hMDAP by using traffic scenarios in a real world.
\end{abstract}


\section{Introduction}\label{sec:introduction}
Data analysis has become a useful technique to organize, process, and analyze large amounts of data in order to obtain useful knowledge effectively such as hidden patterns, implicit correlations, future trends, customer preferences, valuable  business information etc \cite{DAP}. 
OLAP (\emph{online analytical processing}) \cite{OLAP}, as a key technology to provide rapid access to data (mostly relational data) for analysis via multidimensional structures, enables users (e.g., analysts, managers, executives etc.) to gain useful knowledge from data in a fast, consistent, interactive accessing way. 
There are many popular enterprise database management systems for supporting OLAP. For example, Oracle OLAP \cite{OrcaleDW,OrcaleOLAP} is Oracle's current computing engine for online analytical processing. IBM company based on the DB2 database proposes the IBM DB2 OLAP Server \cite{IBM,DB2} which can analyze the relational database quickly and directly. Microsoft also provides SQL Server Analytic Services (SSAS) \cite{MDX,MAS} supporting for OLAP to analyze information, tables, and files scattered across multiple databases. 

The characteristics of big data is not confined to only volume and velocity; it is also referred by the variety, variability and complexity of the data \cite{mckinsey2011big,BI-BD}.  Due to the volume, variety and velocity at which the data grows, it is extremely difficult for organisations to process this data for timely and accurate decisions \cite{BigData-AP1}. 
For this challenge, big data analysis \cite{BigData-AP2} has become a tool to slove the problem.
The primary goal of big data analysis is to help companies make more informed business decisions by enabling data scientists, predictive modelers and other analytics professionals to analyze large volumes of transaction data, as well as other forms of data that may be untapped by conventional business intelligence programs \cite{BigData-AP2}. 
Recently, many techniques have been successfully developped for providing big data analysis in various applications. For example, Oracle Bigdata \cite{OracleBD} builds on Hadoop \cite{Hadoop} through Oracle Direct Connector  connecting Hadoop and Oracle databases. SQL Server 2012 \cite{SQL2012} provides the extension service of OLAP and business intelligence on Hadoop to support big data analysis. IBM SmartCloud provides a Hadoop-based analytical software InfoSphere BigInsights \cite{IBMBD} which can connect with IBM DB2. 
However, those existing techniques of big data analysis are mostly based on OLAP which is not effective to process data in various models (e.g., semi-structure \cite{BigData-AP2}), they do not always bring highly accurate analysis due to the variety and variability of big data in a complicated application--for example, the real-time data on the performance of traffic applications or of mobile applications. Besides, how to process big data analysis efficiently is always an important problem when the scale of big data grows exponentially \cite{Nature}.

In this demonstration, we propose a hybrid framework for big data analysis on Apache Spark \cite{MLlib} (a high-performance computing architecture) which builds on HDFS of Hadoop. The framework features a three-layer data process module and a business process module which controls the former. Within this framework, we can support multi-paradigm data process (i.e., a technical connectivity between various disparate process \cite{paradigm}) in order to improve the accuracy of analysis, where various big data analysis techniques  (incl. OLAP, machine learning, and graph analysis etc.) are interoperated to process the analysis of various applications of big data (incl. data cube \cite{DataCube}, intelligent prediction, and complex network etc.) respectively. Moreover, our proposed framework built on Spark can process large-scale data efficiently. Finally, we implement hMDAP and demonstrate the strength of hMDAP by using traffic scenarios in a real world.

\section{Architecture}\label{sec:architecture}
\begin{figure}
  \centering
  \begin{minipage}[t]{0.95\linewidth}
  \scalebox{1.8}{
  \includegraphics[width=0.5\textwidth]{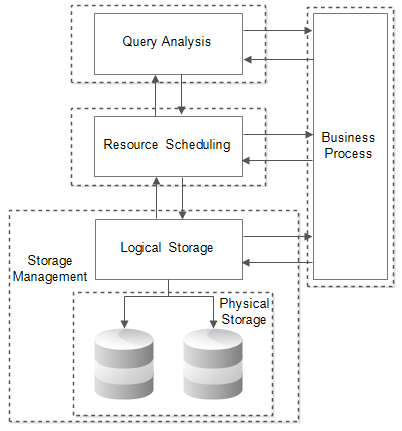}\\
  }
  \caption{The hMDAP architecture.}\label{fig:architecture}
  \end{minipage}
  \vspace*{-10pt}
\end{figure}

In Figure \ref{fig:architecture}, we depict the architecture of our framework consisted of four parts: \emph{the storage management}, \emph{the resource scheduling}, \emph{the query analysis} and \emph{the business process}. In the following sections, we will introduce each part in detail.

\subsection{Storage management}\label{sec:storage}
\begin{figure}[h]
  \centering
  \includegraphics[width=0.5\textwidth]{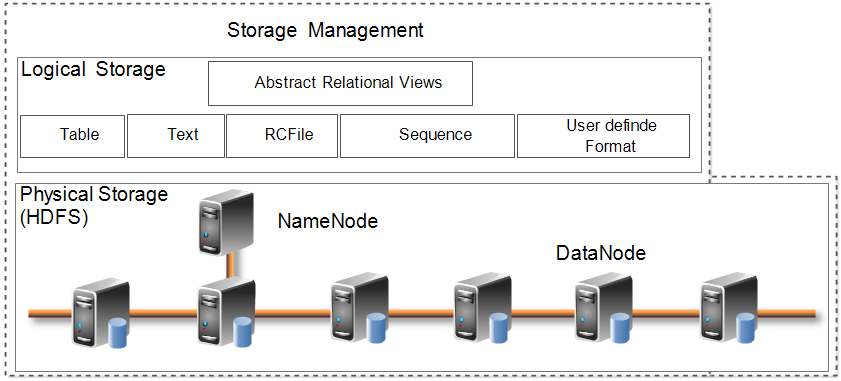}\\
  \caption{Storage management.}\label{fig:storage}
\end{figure}

In Figure \ref{fig:storage}, there are two parts, the physical storage and the logical storage. The rapid growth of data makes the physical storage of data from single source storage to distributed storage. In order to solve the storage of multi-source data, we adopt the existing distributed file system. In our framework, it is HDFS (Hadoop Distributed File System \cite{Hadoop}).

Besides, it products many types of data due to the different needs of applications, such as tables, texts, RCFile(the file type of Hive) and sequence data. In order to use these different types of data, we compose the abstract relational views by designing the metadata with semantics to convert data types to the relational data we can handle.

\vspace*{-15pt}

\subsection{Resource scheduling}\label{sec:resource}
\begin{figure}[h]
  \centering
  \includegraphics[width=0.5\textwidth]{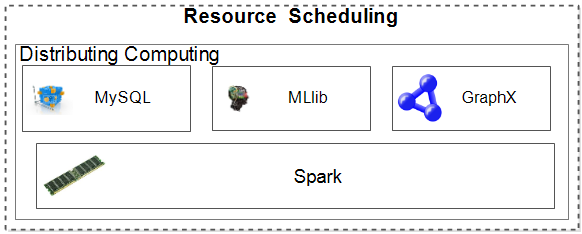}\\
  \vspace*{-10pt}
  \caption{Resource scheduling.}\label{fig:scheduling}
  \vspace*{-20pt}
\end{figure}

In our framework, the development is based on Spark and the module of the resource scheduling is assigned to Spark. The Figure \ref{fig:scheduling} depicts the resource scheduling in our framework. We use MySQL \cite{MySQL} to query over relational database. The part of MLlib is Spark machine learning library. We call the functions in the library to compute. GraphX is the graph query module of Spark. We user it to query graphs and it provides a possibility to transform the different data formats to graph to query.

\subsection{Query analysis}\label{sec:query}
The module of the query and analysis is located on the top of the framework. It is not only the entrance to provide services, but also provides the standard syntax and semantic specification of multi-paradigm data analytical processing. At present, HiveQL is similar to the standard SQL, which is oriented to the classic OLAP task, and does not deal with the query language based on ML analysis and graph data analysis. On the basis of not changing the existing query language syntax standard, we develop a multi paradigm for large data fusion analysis query language expanded of machine learning(ML) and graph analysis.

Our big data analysis and processing of the query language is based on the improvement of the fusion of SQL and HiveQL in multi-paradigm. First of all, we analyze the support of HiveQL and SQL respectively and count the amount of operations which can be supported by the traditional relational algebra model. On the basis of the relational algebra model, we add other necessary operators to construct an extension of the algebraic language model, which can fully support the operation of HiveQL and standard SQL. For the operator with higher complexity, it is split into smaller sub operator or used other methods to optimize it. For the ML analysis, we count the commonly used analytical processing methods, such as classification and clustering, and define the abstract interfaces for common ML analysis processing methods. For the graph analysis processing, we also count the commonly used analytical processing methods, such as the shortest path algorithm, and define the abstract interfaces for them.

In this module, the framework also relates to the implementation of the OLAP on the relational database and ML and graph data processing tasks on the distributed framework. The traditional relational database query optimization method is no longer applicable to this situation. According to the different characteristics of relational storage management query engine and distributed file system of computing engine, we summarize the query information and optimize the performance. Firstly, we investigate the statistical index system used in traditional database and analyze the interaction between each index and the index in the system. Then, for each index in the index system of statistical information, we design efficient and accurate sampling methods to calculate the cost model in query optimization. According to the above statistics, we can also design a storage and maintenance programs which is easy to update and manage. And we may use the cost model in the traditional relational database to design a new cost model which can reflect the query cost of the mixed data.

\begin{figure}[t]
  \centering
  \includegraphics[width=0.5\textwidth]{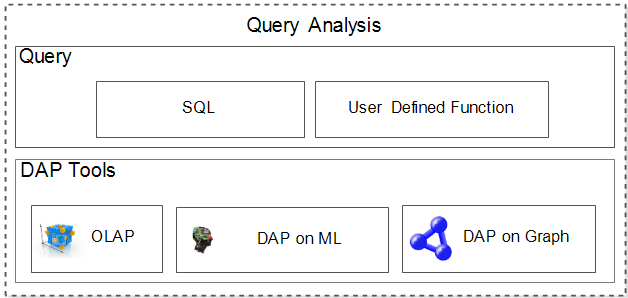}\\
  \vspace*{-10pt}
  \caption{Query analysis.}\label{fig:query}
  \vspace*{-15pt}
\end{figure}

Figure \ref{fig:query} displays the query analysis. The main architectural components of the query analysis are \emph{Query} and \emph{Data Analysis Process Tools(DAP Tools)}. In the first part, we can query by SQL or the function user defined as specified format. The DAP tools contain classical OLAP, DAP on machine learning and DAP on graph.

\subsection{Business process}\label{sec:business}
Our framework provides an analysis method for the large scale data analysis process. But in the face of complex business processes in different fields, we need the domain knowledge and according to the domain knowledge, we can design the multi-paradigm fusion of analysis task. We can draw lessons from the method of service composition in service oriented architecture design.

In this module, we need to do two things: developing a multi paradigm fusion analysis process orchestration language syntax and the complex business process scheduling method. In the first part, we need to analyze the patterns and characteristics of service orchestration language in service oriented architecture design and design an abstract model of the executable process. On the basis of the abstract model, we summarize the basic activities of complex business process analysis. Finally, we define the grammar of the business process. In the semantic, we need to research and analysis the meanings of basic business activities and define the start point,end point and the basic command. In the second part, we need to study and analyze complex business processes in practical applications. Then, we build complex business process models and refine the way to exchange messages in public business processes. After that, we need to control the interaction of each part of the resources through the interaction sequence of messages, achieving a reasonable call for each resource service. We still need to investigate the applicability of existing object-oriented design patterns. For the analysis of complex business process integration model, we design data business processes. We refine the design patterns in complex business processes based on the advantages and principles of existing design patterns.

\begin{figure}
  \centering
  \begin{minipage}[t]{0.95\linewidth}
  \scalebox{1.8}{
  \includegraphics[width=0.5\textwidth]{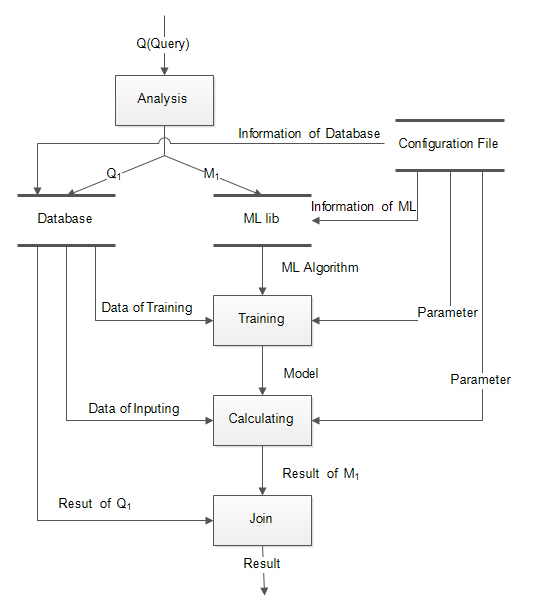}\\
  }
  \vspace*{-15pt}
  \caption{Business process.}\label{fig:business}
\vspace*{-15pt}
  \end{minipage}
\end{figure}

In the real world, the business process model is complex and it takes a lot of time to analyze. The Figure \ref{fig:business} illustrates the details of the business process in our framework.

The user needs to write the configuration files before he or she submits the query. The format of configuration files are shown in Section \ref{sec:demonstration}. When the user submits a query to the framework, the query and analysis module in the framework starts to parse the user's query. This module parses queries according to predefined semantics, such as XML(Extensible Markup Language). The module transforms the user's query to two parts, the query over relational databases and the query in machine learning. We default that the user's query including the query over relational databases and the module determines whether or not to carry out the query in machine learning. We think that when the result of the query over relational databases is null, the framework begins to query in machine learning. After the analysis module, the framework uses the query over relational databases and the information about the databases which is read from the configuration files to query the relational databases. Then, the framework runs the query in machine learning. The input of machine learning is the result of querying by relational databases which the query statement is stored in the configuration files. And the parameters of the machine learning algorithm is also stored in the configuration files. When the framework gets the information of the machine learning algorithm, it starts to train and calculate and the parameters of the training of the machine learning also comes from the configuration files. Finally, the framework makes a join of the results of two parts.
 
\vspace*{-10pt}
\section{Demonstration}\label{sec:demonstration}
\begin{figure}
  \centering
  \begin{minipage}[t]{0.95\linewidth}
  \scalebox{1.8}{
  \includegraphics[width=0.5\textwidth]{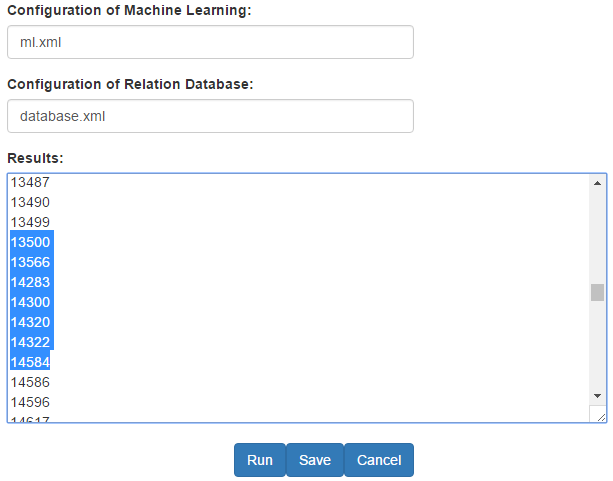}\\
  }
    \vspace*{-10pt}
  \caption{Query interface of hMDAP.}\label{fig:screenshot}
  \end{minipage}
    \vspace*{-10pt}
\end{figure}

\begin{figure}[h]
  \centering
  \begin{minipage}[t]{0.95\linewidth}
  \scalebox{1.8}{
  \includegraphics[width=0.5\textwidth]{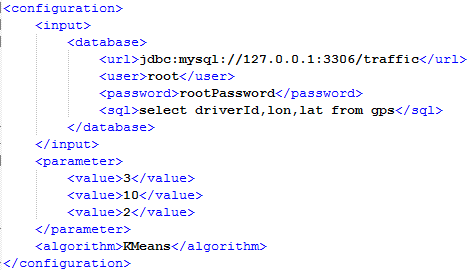}\\
  }
    \vspace*{-10pt}
  \caption{The configuration file of the machine learning.}\label{fig:ml}
  \vspace*{-20pt}
  \end{minipage}
\end{figure}

In this section, we present the interface of hMDAP based in Javascript, which communicate with the service in Java. We show the screenshot of hMDAP in Figure \ref{fig:screenshot} and the configuration file we mentioned above in Figure \ref{fig:ml}.

The interface is composed as follows:
\begin{compactitem}
\item Configuration of Machine Learning: it is a text to input the path of the configuration file of the machine learning algorithm, such as parameters.
\item Configuration of Relation Database: it is a text to input the path of the configuration file of the relational databases, such as the user name.
\item Results: it is a text to display the results of the background.
\item Run: it is a button to start the program and when the program runs over, the results are shown in the \emph{Results}.
\item Save: it is a button to save the context from \emph{Results} in text file and at the same time, empty all the text box contents.
\item Cancel: cancel the running of this program and empty all the text box contents.
\end{compactitem}

And the details of the configuration file is as follows:
\begin{compactitem}
\item configuration: it is the beginning of the configuration file.
\item input: it is the training dataset of the machine learning algorithm.
\item database: it indicates that the input dataset comes from the relational database as following information
\item url, user, password: they are the parameters to connect to the relational database, the location of the database, the user name and the password of the user.
\item sql: it is the statement to query the relational database.
\item parameter: the contents under this label are the parameters of the machine learning algorithms except the input parameter.
\item value: a series of these labels are the values of the parameters.
\item algorithm: it is the name of algorithm. For example, the value of \emph{algorithm} is \emph{KMeans} and our framework runs the algorithm named \emph{KMeans} which is defined in our library. User can customize the algorithm and give the location of the algorithm in this label.
\end{compactitem}

Before running the interface, the user should write two configuration files, the configuration of machine learning algorithms as Figure \ref{fig:ml} and the configuration of relational databases that the contexts are the parts of \emph{<database>} in Figure \ref{fig:ml}. When the user writes two files, he or she should write the paths of the files in the texts on the interface. Then, click the button \emph{Run}. If the user wants to save the results, he or she clicks the button of \emph{Save}. If the user don't need the results, he or she clicks the button of \emph{Cancel}.

\vspace*{-10pt}
\section{Conclusion}\label{sec:conclusion}
In this demonstration, we proposed hMDAP, a hybrid framework for large-scale data analytical processing to support multi-paradigm process on Spark. The  multi-paradigm processing mechanism of hMDAP can provide the interoperability of data analytical process techniques to process data which might be not effectively handled if we only apply single data analytical process technique. On the other hand, hMDAP takes advantage of the high-performance of Spark in processing large-scale data.  We believe that hMDAP provides a new approach to big data analysis in a multi-paradigm way. 

\vspace*{-10pt}
\section*{Acknowledgments}\label{sec:acknowledgments}
This work is supported by the programs of the Key Technology Research and Development Program of Tianjin (16YFZCGX00210), the National Key Research and Development Program of China (2016YFB1000603), the National Natural Science Foundation of China (NSFC) (61672377), and the Open Project of Key Laboratory of Computer Network and Information Integration, Ministry of Education (K93-9-2016-05). Xiaowang Zhang is supported by Tianjin Thousand Young Talents Program.

%

\vspace*{-10pt}
\begin{thebibliography}{10}

\bibitem{BigData-AP1}
Ahmed H. (2015).
\newblock Importance of big data analytics for business growth.
\newblock {\em : BIG Data Analytics News}
\newblock \url{http://bigdataanalyticsnews.com/importance-of-big-data-analytics-for-business-growth/}


\bibitem{DB2}
Baragoin C., Bercianos J., Komel J.,  Robinson G., Sawa R., and 
Schuinder E. (2001).
\newblock DB2 OLAP server theory and practices.
\newblock {\em International Technical Support Organization}.

\bibitem{DAP}
Berson A. and J. Smith S. (1997).
\newblock Data warehousing, data mining, and OLAP.
\newblock {\em McGraw-Hill}.


\bibitem{IBM}
Bontempo C. and Zagelow G.(1998).
\newblock The IBM data warehouse architecture.
\newblock {\em Commun. {ACM}}, 1998, 41(9): 38-48.

\bibitem{Nature}
Campbell P. (editor). (2008).
\newblock Big data: science in the petabyte era.
\newblock {\em Nature}, 455:1–136.

\bibitem{OLAP}
Chaudhuri S. and Dayal U. (1997).
\newblock An overview of data warehousing and {OLAP} technology.
\newblock  {\em {SIGMOD} Record}, 26(1): 65--74.


\bibitem{BI-BD}
Chen H., Chiang R.H.L., and Storey V.C. (2012).
\newblock Business intelligence and analytics: From big data to big Impact.
\newblock {\em MIS quarterl},  36(4):1165--1188.

\bibitem{OrcaleDW}
Dodge G. and Gorman T.(1998).
\newblock Oracle data warehousing.
\newblock John Wiley \& Sons, Inc.

\bibitem{DataCube}
Gray J., Chaudhuri S.,  Bosworth A., Layman A., Reichart D., Venkatrao M., Pellow F., and Pirahesh H. (1997).
\newblock Data cube: A relational aggregation operator generalizing group-by, cross-tab, and sub-totals.
\newblock {\em Data Min. Knowl. Discov.}, 1(1): 29-53.

\bibitem{Hadoop}
Hadoop. (2015).
\newblock \url{http://hadoop.apache.org/}


\bibitem{mckinsey2011big}
Manyika, J., Chui, M., Brown, B., Bughin, J., Dobbs, R.,  Roxburgh, C., and H. Byers, A. (2011).
\newblock Big data: The next frontier for innovation, competition, and productivity.
\newblock {\em McKinsey Global Institute}.


\bibitem{MLlib}
Meng X., K.Bradley J., Yavuz B., R.Sparks E., Venkataraman S., Liu D., Freeman J., B.Tsai D., Amde M., Owen S., Xin D., Xin R., M. Franklin M., Zadeh Z., Zaharia M., and Talwalkar A. (2016).
\newblock MLlib: Machine learning in Apache Spark.
\newblock {\em  J. Mach. Learn. Res.}, 17:1--7.

\bibitem{MAS}
Microsoft Analysis services. (2016).
\newblock SQL server 2016 and later.
\newblock \url{https://msdn.microsoft.com/en-us/library/bb522607.aspx}


\bibitem{OracleBD}
Plunkett T., Macdonald	 B., Nelson	 B., HornickM., Sun	H., Mohiuddin K., Harding	D., Mishra G., Stackowiak R.,  Laker, K., and Segleau	D. (2013). 
\newblock Oracle big data handbook.
\newblock {\em McGraw-Hill Osborne Media}


\bibitem{multi-paradigm}
P. Sheth A., Kochut K., A. Miller J., Worah D.,  Das S.,  Lin C., Palaniswami D.,  Lynch J., and Shevchenko I. (1996).
\newblock Supporting state-wide immunisation tracking using multi-paradigm workflow technology.
\newblock In: {\em Proc. of VLDB'96}, pp. 263--273.

\bibitem{BigData-AP2}
Rouse W. (2012).
\newblock What is big data analytics?
\newblock {\em TechTarget.com}
\newblock \url{http://searchbusinessanalytics.techtarget.com/definition/big-data-analytics}


\bibitem{SQL2012}
Sarkar D. (2013).
\newblock Microsoft SQL server 2012 with Hadoop.
\newblock {\em Packt Publishing}


\bibitem{OrcaleOLAP}
Schrader M. and Vlamis D.(2009).
\newblock Oracle Essbase \& Oracle OLAP.
\newblock {\em Peter Gbolagade Akintunde}.

\bibitem{MDX}
Spofford G. (2001).
\newblock MDX solutions: with Microsoft SQL server analysis services.
\newblock {\em Wiley}.


\bibitem{IBMBD}
Zikopoulos P. (2011).
\newblock Understanding big data: Analytics for enterprise class Hadoop and streaming data.
\newblock {\em McGraw-Hill Osborne Media}


\bibitem{paradigm}
zur Muehlen M. and Rosemann M.(2004).
\newblock  Multi-paradigm process management. 
\newblock In: {\em Proc. of CAiSE'04}, pp. 169--175.

\end{thebibliography}
%
%

\end{document}